\documentclass[reprint, aps, prl,showpacs,superscriptaddress]{revtex4-1}
\usepackage{graphicx}
\usepackage{amsmath}
\usepackage{hyperref}

\usepackage{txfonts}

\newcommand{\Imax}{\langle I_\mathrm{S,max}\rangle}
\newcommand{\Iav}{\langle I_\mathrm{S}\rangle}
\newcommand{\Ips}{I_\mathrm{PS}}
\newcommand{\Ir}{I_\mathrm{R}}

\newcommand{\muin}{\mu_0}
\newcommand{\dmu}{\Delta \mu}
\newcommand{\dmuF}{\Delta \mu_\mathrm{F}}
\newcommand{\Nr}{N_{\mathrm{D}}}

\newcommand{\vB}{v_{\mathrm{WL}}}

\newcommand{\Vr}{U_{\mathrm{WL}}}

\newcommand{\nin}{n_\mathrm{1D}}
\newcommand{\xiin}{\xi_0}

\newcommand{\nB}{n_\mathrm{WL}}
\newcommand{\xiB}{\xi_\mathrm{WL}}

\newcommand{\dphi}{\Delta \phi}
\newcommand{\phic}{\Delta \phi_c}

\newcommand{\Nexc}{N_\mathbf{exc}}

\newcommand{\OurNa}{~^{23}\mathrm{Na}}
\newcommand{\Hz}{\mbox{~Hz}}
\newcommand{\kHz}{\mbox{~kHz}}
\newcommand{\nK}{\mbox{~nK}}

\newcommand{\ms}{\mbox{~ms}}
\newcommand{\s}{\mbox{~s}}

\newcommand{\degAng}{~^\circ}
\newcommand{\ums}{~\mu\mathrm{m}/\mathrm{s}}

\newcommand{\mum}{~\mu\mathrm{m}}

\hypersetup{
    colorlinks=true,       
    linkcolor=blue,          
    citecolor=blue,        
    filecolor=blue,      
    urlcolor=blue           
}

\begin{document}

\title{Resistive flow in a weakly interacting Bose-Einstein condensate}
\author{F. Jendrzejewski}
\author{S. Eckel}
\affiliation{Joint Quantum Institute, National Institute of Standards and Technology and University of Maryland, Gaithersburg, Maryland 20899, USA. }
\author{N. Murray}
\author{C. Lanier}
\author{M. Edwards}
\affiliation{Department of Physics, Georgia Southern University, Statesboro, Georgia 30460-8031, USA.}
\author{C. J. Lobb}
\author{G. K. Campbell}
\affiliation{Joint Quantum Institute, National Institute of Standards and Technology and University of Maryland, Gaithersburg, Maryland 20899, USA. }


\begin{abstract}
We report the direct observation of resistive flow through a weak link in a weakly interacting atomic Bose-Einstein condensate (BEC). Two weak links separate our ring-shaped superfluid atomtronic circuit into two distinct regions, a source and a drain. Motion of these weak links allows for creation of controlled flow between the source and the drain. At a critical value of the weak link velocity, we observe a transition from superfluid flow to superfluid plus resistive flow. Working in the hydrodynamic limit, we observe a conductivity that is four orders of magnitude larger than previously reported conductivities for a BEC with a tunnel junction. Good agreement with zero-temperature Gross-Pitaevskii simulations and a phenomenological model based on phase slips indicate that the creation of excitations plays an important role in the resulting conductivity. Our measurements of resistive flow elucidate the microscopic origin of the dissipation and pave the way for more complex atomtronic devices.
\end{abstract}

\maketitle
 Resistivity, and hence dissipation, plays an important role in the behavior of many superfluid and superconducting systems and is essential in the operation of devices like dc superconducting quantum interference devices (SQUIDs) \cite{clarke2004squid, Sato2012}. Such dissipation occurs above a critical velocity in superfluids, as first observed in liquid helium \cite{Allum1977,Avenel1985}. Degenerate quantum gases of neutral atoms \cite{Bloch2008} and of polaritons  have provided new possibilities for studying the superfluid state \cite{Carusotto2013}. Motion of a perturbing potential in an atomic Bose-Einstein condensate (BEC) provided evidence for a critical velocity \cite{Raman1999,Onofrio2000, Desbuquois2012} and allowed for observing the onset of excitations like vortices and solitons \cite{Inouye2001,Engels2007,Neely2010,Ramanathan2011a, Wright2013a}. 
 
Recent experiments, where a Fermi gas discharges from a source region into a drain, allowed for the observation of the drop in resistance across the superfluid transition \cite{Stadler2012}. A similiar experiment studied the resistive transport of a thermal Bose gas through a channel \cite{Lee2013}. In contrast, here we control the current of a weakly interacting BEC between two regions, a source and a drain, and measure the resulting time evolution of the chemical potential difference. We show that the BEC exhibits both superfluid and resistive flow through a constriction, i.e., a weak link \cite{Likharev1979}. We describe our observations using a phenomenological model incorporating dissipation through excitations from the superfluid ground state and find good agreement with simulations using the three-dimensional, zero-temperture Gross-Pitaevskii equation (GPE).   Such transport measurements could be extended to the study of a wide variety of exotic quantum matter providing insight into the emergence of resistive flow in the absence of a thermal component. Moreover, they pave the way for creating more complex atomtronic devices \cite{Pepino2009} such as an atomic dc SQUID \cite{Ryu2013a}.
 
 \begin{figure}[!ht]
\begin{center}
\includegraphics[width=1\linewidth]{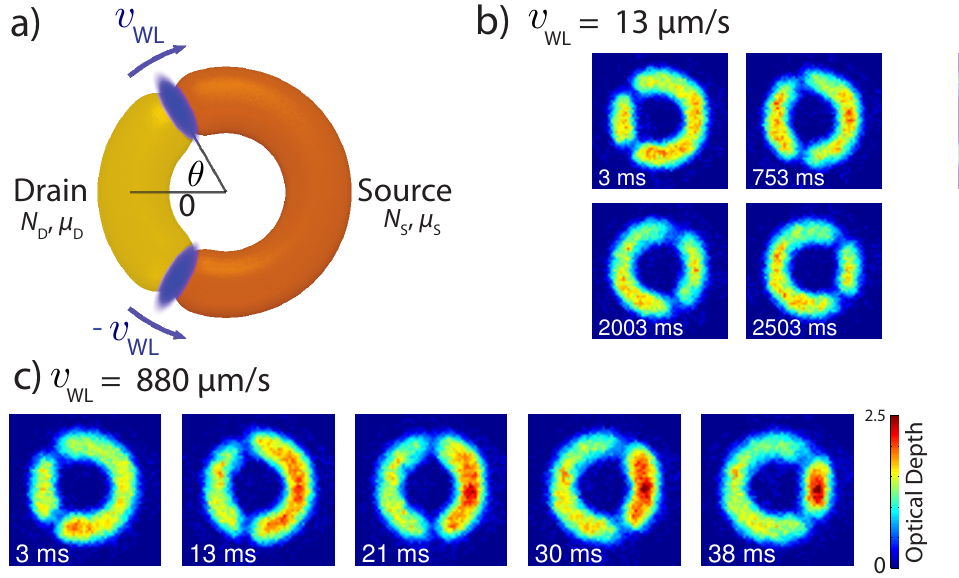}
\caption{The experimental setup. a) Two repulsive potentials at $\pm \theta$ form two weak links and separate two regions, i.e., a source and a drain. Both weak links are set into motion with equal and opposite speed $\vB$, inducing a flow of atoms from the source into the drain. The regions are characterized by the number of atoms $N_{S,D}$ and the chemical potential $\mu_{S,D}$. b)   \textit{In-situ} absorption images of the atomic density after a variable time $t$. For this speed ($\vB = 13\ums$), we observe no change in the density far from the barriers. c) On the other hand,  a barrier speed of $\vB = 880\ums$ results in a density difference, which corresponds to a chemical potential difference that allows for resistive flow. In b,c) the strength of the weak link is $\Vr/\muin = 0.53(6)$, where $\muin$ is the chemical potential before applying the weak links. The field of view in b) and c) is $72\mum\times72\mum$.}
\label{fig:Experiment}
\end{center}
\end{figure}

Our experiment uses a BEC of $\OurNa$ atoms in a ring-shaped optical dipole trap, as shown in Fig.~\ref{fig:Experiment}(a). A blue-detuned laser creates two repulsive potentials, i.e. two weak links, and separates the ring into two distinct regions. The phase of the condensate wave function can be different between the two regions, and such a phase difference, $\dphi$, results in a superfluid current. Like any superfluid current, it can exist even without a chemical potential difference between the two regions. To drive such a current, we move the two weak links towards each other at speed $\vB$. This induces a flow of atoms, $I = \frac{d\Nr}{dt}$, from the source into the drain (Fig.~\ref{fig:Experiment}(a)), where $\Nr$ is the number of atoms in the drain. 
 The motion of the two weak links changes the volume of the two regions and, for small enough $\vB$, induces a time-averaged superfluid current $\langle I_S\rangle = 2 \nin\vB$.  Here, $\nin$ is the effective linear density, which is the same in the two regions. Fig.~\ref{fig:Experiment}(b) shows that, in this regime, the atomic density far from the weak link stays constant in time.  

In contrast, for larger velocities, see Fig.~\ref{fig:Experiment}(c), a density difference between the source and the drain develops over time. As $\vB$ is increased,  a critical phase difference $\phic$ is reached. The simplest model, a superflow model, assumes that the total, superfluid, current through the weak link cannot exceed a maximum value $\Imax$, which occurs when $\phic$ is reached.  If $\vB$ is increased beyond this critical point, the increasing density in the source causes an increasing chemical potential in this region. \footnote{In these experiments we drive the system out of equilibrium, but each reservoir should independently remain in equilibrium.}
 Using the superflow model, this increasing chemical potential results from compression, as $\Imax < 2\nin \vB$. This approach has been successfully used to describe several ultracold atom experiments that studied flow through tunnel junctions \cite{Albiez2005,Ryu2013a}. In this pure superflow model, only the existence of an additional thermal 
 component can give rise to resistive flow \cite{Levy2007}.

Above the critical phase difference the superfluid current can become unstable, an effect not included in the superflow model. These instabilities can lead to phase slips in the moving weak link region, resulting in excitations like solitons or vortices which are shed into the drain \cite{Anderson1966,Avenel1985,Wright2013}. Such excitations are associated with an additional current. This current is dissipative, i.e. resistive; the excitations eventually decay. Given this additional resistive current, the total  current between regions is larger than in the superflow model and hence a smaller chemical potential difference between the two regions occurs. To sustain the resistive current, a chemical potential difference $\dmu(t)$ between the source and the drain must be present \cite{tilley1990superfluidity}. 

A ring-shaped, all-optical dipole trap confines the BEC, which contains $\approx 7.5\times10^5$ atoms, as described in Ref. \cite{Eckel2014}. The trap frequency is $512(4) \Hz$ in the vertical direction \footnote{Unless stated otherwise, uncertainties are the uncorrelated combination of $1\sigma$ statistical and systematic uncertainties.}. The toroid has mean radius of $R = 20.0(4) \mum$ and a radial trap frequency of $\approx 260$ Hz. The initial chemical potential is $\muin/h \approx 3 \kHz$, where $h$ is the Planck constant.    The atomic density profile is well described by a Thomas-Fermi profile, without a discernible thermal component. We estimate that the temperature is $\leq 0.5 T_C\simeq 300\nK$ and the condensed fraction is $\geq80\%$.

After creating the ring-shaped BEC, we linearly ramp the potential height of the weak links to their final value, $\Vr$, in $200\ms$, while keeping their position fixed at $\theta = \pm 45\degAng$. The weak link is created by a optical, repulsive potential that is homogeneous in the radial direction and has a $1/e^2$ half-width of $\approx5.9 (4) \mum$ in the azimuthal direction \cite{Eckel2014}. The weak link is thus much wider than the healing length of the condensate ($\xiin = \hbar/\sqrt{2m\muin}\simeq 0.27\mum$)
 , and therefore we treat the density within the barrier using the local density approximation (LDA). The density profile in the weak link region enables the calibration of the potential height,  as described in Ref. \cite{Eckel2014}.  Next, the weak links are suddenly set into motion with constant $\vB$. To measure the chemical potential difference across the weak links, we use \textit{in-situ} partial-transfer absorption imaging \cite{Ramanathan2012}, which yields the atomic density distribution of the whole condensate. The half-width at $1/e^2$ imaging resolution is $3.4\mum$.

\begin{figure}[!ht]
\begin{center}
\includegraphics[width=1\linewidth]{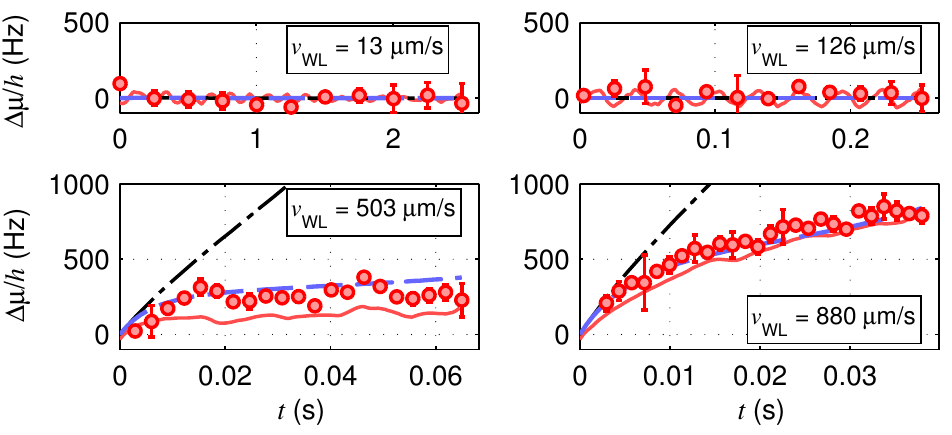}
\caption{Time evolution of the chemical potential difference $\dmu$ as function of time for different weak link speeds $\vB$. For this data, the strength of the weak link is $\Vr/\muin = 0.53(6)$, where $\muin/h = 3\kHz$. The red circles represent the experimental data. The black dash-dotted line shows the superflow model, which does not accurately predict the increase in $\dmu$ (lower row). In contrast, the data show good agreement with the resistive flow model (blue dashed line). The GPE calculation (red line) shows qualitative agreement.}
\label{fig:Tevolv}
\end{center}
\end{figure}

Fig.~\ref{fig:Tevolv} shows the time evolution of $\dmu $ for a range of $\vB = 13\ums$ to $880\ums$ at a weak link strength of $\Vr/\muin = 0.53(6)$, where $\dmu = \mu_S-\mu_D$ is calculated from the chemical potential in the source $\mu_S$ and  drain $\mu_D$. Both $\mu_{S}$ and $\mu_D$ are obtained from the  the measured density. For small speeds (upper panel), the chemical potential difference stays constant around zero. For larger speeds (lower panel), the final chemical potential difference increases monotonically. 
 
\begin{figure}[!ht]
\begin{center}
\includegraphics[width=1\linewidth]{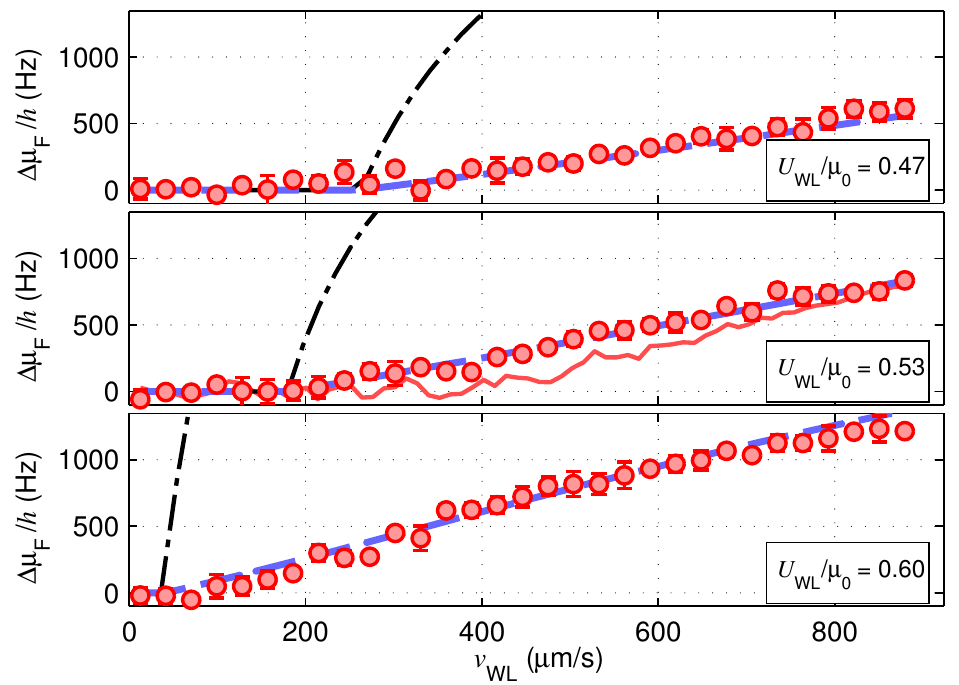}
\caption{Final chemical potential difference for different weak link speeds and strengths. $\dmuF$ is measured  after a sweep angle of approximately $90\degAng$. Shown are the experimental data (red circles), the flow model incorporating resistance (blue dashed line) and the superflow model (black dash-dotted line). The results of the GPE calculation (red line) are compared to the experimental data in the middle panel. The middle panel corresponds to the potential strength presented in Fig.~\ref{fig:Experiment} and \ref{fig:Tevolv}.}
\label{fig:Muevolv}
\end{center}
\end{figure}

The data in Fig.~\ref{fig:Tevolv} are first compared to the superflow model. Here, we assume that the superfluid current $\Iav$ equilibrates the chemical potential in the source and the drain, up to  a maximum value. The time evolution of $\dmu$ is calculated iteratively. For each time step $\Delta t$, the number of atoms in each region $N_{D(S)}(t+\Delta t) = N_{D(S)}(t)\pm \Iav\cdot\Delta t$ and the associated $\dmu(t+\Delta t)$  is determined. Fig.~\ref{fig:Tevolv} presents the result of the calculation for $\Imax = 2.15 \times 10^6~\s^{-1}$  (black dash dotted line) \footnote{This value represents a best fit of Eq. \eqref{Eq:ToyModel} to the data presented in Fig.~\ref{fig:Muevolv}}. The superflow model reproduces the low velocity region 
and, for larger velocities, predicts the short time evolution of $\dmu$. However, it strongly overestimates $\dmu$ for longer times. This suggests that excitations drive an additional, resistive current. 
 
To describe this \textit{resistive regime}, we assume an additional current that obeys the Ohm's law $\Ir = G\dmu$. Thus, the total current is modeled as
\begin{equation}\label{Eq:ToyModel}
I(t) = \Imax + G\cdot\dmu(t)\mbox{ ,}
\end{equation}
with two parameters, $\Imax$ and the conductance $G$. This model (dashed blue lines) agrees well with the data for $\Imax = 2.15 \times 10^6~\s^{-1}$ (as above) and $G= 11.7 \times 10^3/h$. For short times, the increase in the chemical potential difference is mainly due to a compression of the source region. Once the compression establishes a sufficient chemical potential difference, the resistive current becomes crucial for a quantitative description of the experimental data.
  
Additionally, we simulate our experiments using the three-dimensional, time-dependent GPE. The density distributions obtained from the simulations are analyzed in the same way as the experimental data. The time evolution of the resulting chemical potential difference is plotted in Fig.~\ref{fig:Tevolv} (solid red line). The simulations are generally in good agreement, but fail to predict the long time behavior at intermediate weak link speeds.

For small speeds, the simulations show small amplitude oscillations of the chemical potential difference around zero. These oscillations are created during the non-adiabatic change in speed when we set the weak links into motion. This sudden jump creates a density wave which moves around the ring and giving rise to the observed oscillations. In the context of superfluid flow through Josephson junctions, these are known as plasma oscillations and have been observed experimentally in ultra-cold atoms experiments \cite{Albiez2005,Levy2007}. These oscillations represent a $1~\%$ change in the chemical potential  and cannot be detected in this experiment. 

For larger $\vB$, the simulations show an increase in the chemical potential difference that is qualitatively similar to the experiment. In the simulations, this increase in $\dmu$ coincides with the appearance of excitations. This agrees with our assumption that excitations are created above the critical phase difference and are the origin of the resistive component in Eq. \eqref{Eq:ToyModel}. However, their size (on the order of $\xiin$) is below the imaging resolution and thus, they are not directly observable in the experiment. 

\begin{figure}[!ht]
\begin{center}
\includegraphics[width=1\linewidth]{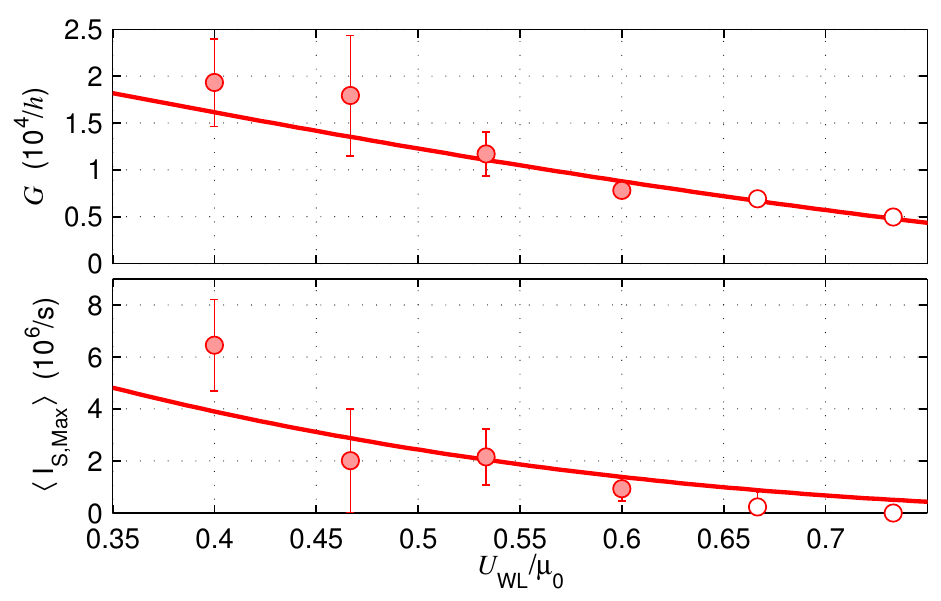}
\caption{Conductance (upper panel) and maximum superfluid current (lower panel) for different weak link strengths. These values (filled circles) are obtained from fits of the data shown in Fig.~\ref{fig:Muevolv} using Eq. \eqref{Eq:ToyModel}. For the highest weak link strengths (open circles), this model fails to describe the data well. We fit the values shown with the estimates of the conductance and the superfluid current obtained from the phase-slip picture (solid red line).}
\label{fig:Overview}
\end{center}
\end{figure}

To identify more clearly the onset of resistive current, in Fig.~\ref{fig:Muevolv} we plot the final observed chemical potential difference $\dmuF$ (red circles)  as a function of the weak link speed, $\vB$. The chemical potential difference is measured after the weak links move by a fixed angle of $\approx 90 \degAng$. We observe a clear separation between the superfluid and resistive regimes. The superfluid regime corresponds to small speeds, at which the chemical potential difference stays constant around zero. If $\vB$ is increased beyond a critical point, the observed final chemical potential difference increases monotonically for larger speeds.  The superflow model (black dash dotted line) shows the onset of a $\dmuF$, but it grossly overestimates its amplitude in this resistive regime. Adding the resistive component to the flow (blue dashed line) lowers the predicted amplitude of $\dmuF$ and enables us to fit the experimental data in both regimes with Eq. \eqref{Eq:ToyModel} \footnote{These uncertainties and those in Fig.~\ref{fig:Overview} are the proper combination of the correlated and uncorrelated $1\sigma$ statistical uncertainties.}. The obtained fit values for $\muin/\Vr = 0.53$ were used in Fig.~\ref{fig:Tevolv}. 

The final chemical potential difference, $\dmuF$, appears to be a continuous function at the onset of resistive current in the experiment and the GPE simulation. In contrast, a jump is predicted to occur for a tunnel junction \cite{Mccumber1968}. Therefore, in contrast to  Refs. \cite{Albiez2005,Levy2007,Ryu2013a} we do not use a tunnel junction model to describe our experiments. Such experiments were mostly described by a pure superfluid flow model. Ref. \cite{Levy2007} introduced a conductance due to the thermal (non-condensed) component in order to explain the observed relaxation of the system. The conductance extracted from Ref. \cite{Levy2007} is 4 orders of magnitude smaller than the conductance observed here.

Comparing the experimental results to 
 GPE simulations for $\Vr/\muin = 0.53$ reveals quantitative agreement both above and below the onset of resistive current (Fig.~\ref{fig:Muevolv}).  However, the GPE simulations and the experiment slightly disagree around the onset of the resistive regime. The simulations continue to show oscillations around zero, while the experiments measure an increase in the chemical potential difference in this regime. It may be that damping of plasma oscillations, absent from the GPE simulations, leads to this discrepancy. 

To gain further insight into the dissipation, we analyze the fitted values of the conductance and maximum superfluid current for different weak link strengths, as shown in Fig.~\ref{fig:Overview}.  At all weak link strengths, we observe both the  superfluid and resistive regime, and with increasing weak link strength, the onset of the resistive regime occurs at smaller speeds. This observation is in qualitative agreement with the numerical simulations.
For each weak link strength $\Vr$, we fit the final chemical potential $\dmuF$ at different speeds to Eq. \eqref{Eq:ToyModel} and extract $G$ and $\Imax$. This model agrees well with the experimental data for $\Vr/\muin\leq 0.6$ (filled red circles). For higher weak link strength, we do not find a good fit (open red circles).

These observations can be compared to estimates of the conductance and the maximum superfluid current. As suggested by the agreement of the experimental results with the GPE simulations, we first estimate the value of the conductance using a phase-slip picture as presented in \cite{Avenel1985}. The relative phase between the two regions follows $\partial_t\dphi = \dmu/\hbar$ \cite{Avenel1985, Dalfovo1999a}, which we integrate, assuming a constant $\dmu$, in a short time step. From this, we estimate the rate of phase slips to be $\frac{\dmu}{\phic\hbar}$, where $\phic$ is the critical phase difference. 
If we further assume that each phase slip creates an excitation that contains $\Nexc$ atoms, we again obtain Ohm's Law with a conductance $G_\mathrm{PS} = \frac{4\pi \Nexc}{\phic h}$. As the healing length in the weak link region  $\xiB$  provides the typical length scale of an excitation, we estimate the number of atoms in each excitation to be $\Nexc = \nB\xiB$, where $\nB$ is the linear density in the weak link region.  Because the critical phase difference is on the order of $\pi$, we estimate the conductance due to phase slips to be on the order of $G_\mathrm{PS} = \frac{4 \nB \xiB}{h}$.  Both $\nB$ and $\xiB$  are calculated in the Thomas-Fermi approximation. We assume that $G=CG_\mathrm{PS}$, where $C$ is a constant whose precise value depends on the type of excitations that give rise to the phase slips. We obtain a good fit to the data with  $C= 5.4(7)$, as shown in the upper panel of Fig.~\ref{fig:Overview}. This agreement provides further evidence that phase slips are relevant for describing the emergence of resistance in our experiments.

In addition to phase slips, non-condensed atoms in the thermal component could contribute to resistive flow. We estimated this contribution from a model where resistive flow arises from ballistic transport of such thermal atoms through the weak link \cite{Lee2013}. Here, we obtain an upper limit for the conductance of the thermal atoms 
 $G_\mathrm{Th} \approx \frac{4n_\mathrm{Th}\lambda_\mathrm{dB}}{h}$, where $n_\mathrm{Th}$ is the density of thermal atoms and $\lambda_\mathrm{dB}$ is the de-Broglie wavelength. This upper bound on $G_\mathrm{Th}$ is of the same order of magnitude as $G_\mathrm{PS}$. This implies that $G_\mathrm{Th}$  could also play a role. However, the good agreement between the experimental data and the GPE simulations suggests that phase slips are important for an accurate description of the observed flow. This interplay between phase slips and thermal atoms will be the focus of future studies.

 The maximum average superfluid current should be smaller than the critical current as the superfluid flow has to be reestablished after each phase slip. One might assume that the  local speed of sound in the barrier region \cite{Watanabe2009} sets the critical current $\Ips = 2\nin c_{s}$. Our measured values of the maximum superfluid current are best fit by $\Imax = 0.8(3)\times\Ips$ \footnote{In this fit, we exclude the maximum currents of barrier heights $\Vr/\muin\geq 0.67$.}, as shown in the lower panel of Fig.~\ref{fig:Overview}. This indicates that the critical current is set by the speed of sound \cite{Piazza2010}, which differs from previous experiments \cite{Raman1999, Onofrio2000, Desbuquois2012, Eckel2014}. This includes experiments with a single weak link \cite{Ramanathan2011a,Eckel2014}, where the critical velocity for phase slips was inconsistent with the speed of sound. We speculate that this discrepancy arises because with a single weak link each phase slip reduced the relative velocity between the barrier and the fluid, whereas here it does not.
  
 In conclusion, we have observed resistive flow of an atomic gas superfluid above a critical current and have measured the conductance. Our results provide evidence that phase slips in the superfluid component play an important role in the observed conductance. This connection between resistive flow and phase slips provides valuable information: in particular, it connects dissipation with excitations from the ground state of the superfluid, without invoking the existence of a thermal component. Such measurements represent a useful and versatile tool that can shed light on the microscopic mechanisms that give rise to resistive flow. Similar conductivity measurements will enable a quantitative study of the out-of-equilibrium properties of fascinating, yet not fully understood, quantum materials like low dimensional Bose gases \cite{Desbuquois2012, Kinoshita2004, Paredes2004}, unitary Fermi gases \cite{Yefsah2013}, dipolar gases \cite{Baranov2012} or superfluids with spin-orbit coupling\cite{Lin2011f}.
 
Our system, which is composed of a ring with two weak links, is reminiscent of a dc SQUID geometry, and thus may appear to be the analogous atomtronic rotation sensor. However, our set-up does not incorporate  current leads, which would provide the necessary splitter and recombiner in an interferometer. This raises a fundamental question: Can quantum interferences be observed in this device without such leads?
\begin{acknowledgments}
This work was partially supported by ONR, the ARO atomtronics MURI, and the NSF through the PFC$@$JQI, grant PHY-0822671, and by PHY-1068761. S.E. is supported by a National Research Council postdoctoral fellowship. We wish to thank W. D Phillips, A. Ran\c{c}on, J. Lee, and A. Kumar for valuable discussions and experimental assistance.
\end{acknowledgments}
%
\end{document}